\documentclass[12pt]{article}

\makeatletter
\@ifclassloaded{apa7}{\newif\ifarxiv\arxivfalse}{\newif\ifarxiv\arxivtrue}
\makeatother

\usepackage{xcolor}
\usepackage{ifthen}
\usepackage[T1]{fontenc}
\usepackage[utf8]{inputenc}
\usepackage[american]{babel}
\usepackage{csquotes}

\usepackage{fancyhdr}
\usepackage{setspace}
\usepackage{float}

\usepackage{amsmath}
\usepackage{amssymb}
\usepackage{amsthm}
\usepackage{amscd}

\usepackage{enumerate}
\usepackage{verbatim}
\usepackage{listings}
\usepackage{endnotes}

\usepackage{multirow}
\usepackage{graphicx}
\usepackage{rotating}
\usepackage{subfig}
\usepackage{wrapfig}

\usepackage{url, multicol}

\usepackage{appendix}

\ifarxiv
  \usepackage{arxiv}
  \renewcommand{\headeright}{}
  \renewcommand{\undertitle}{}
  \renewcommand{\shorttitle}{SPICE}
  \usepackage{hyperref}
  \usepackage{orcidlink}
  \usepackage{authblk}
\fi

\usepackage[style=apa, backend=biber]{biblatex}
\usepackage{detshortcuts}

\let\footnote=\endnote

\def\I{\text{I}}
\def\J{\text{J}}
\def\U{\text{U}}

\DeclareMathOperator{\vect}{vec}
\DeclareMathOperator{\logit}{logit}

\title{A Scalable Parametric Item Calibration Engine (SPICE) for Explanatory IRT with Sparse Data}
\ifarxiv
  \author{Steven W. Nydick \orcidlink{0000-0002-2908-1188}}
  \author{Manqian Liao \orcidlink{0000-0002-8444-9440}}
  \author{J.R. Lockwood \orcidlink{0000-0001-9116-4219}}
  \affil{Duolingo}
  \date{\today}
\else
  \shorttitle{SPICE}
  \authorsnames{Steven W. Nydick, Manqian Liao, J.R. Lockwood}
  \authorsaffiliations{{Duolingo} \\\today}
\fi

\newcommand{\abstracttext}{We describe a Bayesian multidimensional explanatory IRT model, and an associated Markov Chain Monte Carlo (MCMC) estimation procedure and the corresponding development of calibration software, designed for psychometric analyses of large numbers of sparsely-linked persons and items. Such data structures can arise, for example, from adaptive assessments using large banks of automatically generated items with individual test takers receiving a very small proportion of the entire bank. We discuss how our choices for model specification, data structures, and algorithm implementation combine to create a scalable method for explanatory IRT that can support a variety of psychometric operations with sparse data.}

\ifarxiv\else
  \abstract{\abstracttext}
\fi

\begin{document}

\maketitle

\ifarxiv
  \begin{abstract}\abstracttext\end{abstract}
\fi

\section{Background}

Rapid technological advances, including generative AI, are increasing the demand for conducting psychometric analyses with datasets of unprecedented scale. For example, assessment or pedagogical contexts are arising in which tasks from large banks of automatically generated items are adaptively presented to persons. Psychometric analyses of the resulting data structures may require methods that can handle large numbers of sparsely-linked persons and items and multidimensional latent variables. Given the small sample sizes per item, these methods might also benefit from feature vectors to explain population variation of those latent variables and reduce the sample size necessary for adequately recovering item parameters \parencite[e.g.,][]{mccarthy2021, yancey2024}. The purpose of this paper is to specify such a multidimensional explanatory item response theory (IRT) model, as well as the Bayesian theory and associated Markov Chain Monte Carlo estimation procedure, designed for such circumstances.

Our methods are optimized for sparse linkage structures, in which there are large numbers of two (or more) distinct types of things but relatively few observed pairings. Sparse linkage structures arise in measurement applications, such as matrix-sampled assessments \parencite{mislevy:etal:1992} and computerized adaptive tests (CAT) \parencite{chang:2015,vanderlinden:chang:2020,weiss1982}, in which each test taker is exposed to only a small fraction of items. Sparse linkage structures also occur in a variety of psychological data, such as in the analysis of worker and firm inputs to productivity \parencite{Abow:Kram:Marg:1999}, the analysis of teacher, school, and student inputs to student achievement \parencite{Hanu:1979,McCa:Lock:Kore:Loui:Hami:2004,raudenbush:2004,Rivk:Hanu:Kain:2005}, and other contexts involving complex relationships among analysis entities \parencite{Brow:Gold:Rasb:2001,Gold:1994,huang:cai:2024,rohloff:etal:2024}. Analysis methods using dense representations of linkage structures become impractical when the number of potential pairings grows, motivating procedures designed to accommodate sparsity
\parencite{Andr:Scha:Upwa:2006,Clay:Rasb:1999,Lock:McCa:Mari:Seto:2007,rpackage:lme4}. 

Two general aspects of the model are worth highlighting. First, both persons and items are modeled with multidimensional latent variables (i.e., person parameters/traits and item parameters). Allowing for multidimensionality on the person side accommodates multi-stage tests or multidimensional CATs, for which joint analysis of all item responses is generally required for unconfounded estimation of item and person parameters \parencite{jewsbury:vanrijn:2020}. Allowing for multidimensionality on the item side supports calibration of any finite-dimensional parametric IRT model that treats item parameters as random vectors \parencite{deboeck:2008}. For both persons and items, latent variable vectors are modeled with multivariate regressions on observed features, analogous to those used in some large-scale assessments \parencite{delatorre:2009,johnson:jenkins:2005,mislevy:etal:1992,vondavier:sinharay:2010} and item explanatory modeling \parencite{deboeck:wilson:2004}. Our latent regression implementation accommodates high-dimensional feature sets via optimized linear algebra routines.

Second, this model has been implemented in a Bayesian framework. Bayesian approaches to item calibration have precedent for IRT calibration \parencite[e.g.,][]{albert:chib:1993,deboeck:wilson:2004,fox:glas:2001,fox:2010,levy:mislevy:2017,patz:junker:1999} and continue to be an active area of research \parencite{burkner:2021,konig:etal:2020,lin:etal:2024,luo2026,merkle:rosseel:2018,ulitzsch:etal:2026,wu:etal:2020,xia:etal:2026,xu:etal:2024}. Moreover, even though our model could be specified in a general Bayesian modeling language, such as Stan \parencite{stan:2015}, we opted for a custom implementation to support scalability to the large numbers of persons, items, and features demanded by modern computational psychometric contexts \parencite{avd:mislevy:hao:2021} as well as to avoid an inefficient translation to the underlying algorithm possible in general purpose Bayesian estimation software. As described by \textcite{sharpnack2026}, 
the methods described here are suitable for periodic calibration of items used in adaptive assessment systems with rapid item turnover supported by automatic item generation.

In the following sections, we specify the statistical model for the observed and latent variables. We then present the estimation algorithm and conclude with a discussion of the strengths and limitations of the model and directions for future developments. Longer derivations are provided in the Appendix.

% insert "followed by a demonstration using simulated data" in the published paper!

\section{Bayesian Statistical Model}

We begin by establishing terminology and notation for persons, items, and their linkages. We then specify models for latent variables and item responses. We finally indicate prior distributions and identification constraints.

\subsection{Terminology and Notation for Linkage Structures}

Let both persons and items be associated with latent variables. For persons, latent variables equate to proficiency dimensions, denoted by $\boldsymbol{\theta}$. For items, latent variables are item model parameters for a given IRT model, denoted by $\boldsymbol{\psi}$. For instance, if $\boldsymbol{\theta}$ is one-dimensional, then $\boldsymbol{\theta} = (\theta)$ represents the proficiency dimension vector (i.e., the latent variables on the single person dimension). Moreover, if responses to a set of items can be modeled with a two-parameter logistic (2PL) item response function, then $\boldsymbol{\psi} = (d, a)$ represents the item model parameters (i.e., the latent variables on the item dimension). In this example, the person latent variable is represented by a one-dimensional vector whereas the item latent variable is represented by a two-dimensional vector. In the remainder of this document, we use the term \emph{units} to refer to either persons or items.

Units are partitioned into mutually exclusive subpopulations, referred to as \emph{blocks}. Units in the same block share a common probability model for their corresponding latent variables (specified in a later subsection). The numbers of person and item blocks depends on the data and objectives of any given application of the model. For example, calibrating a collection of unidimensional 2PL items with respect to a single person population may group all persons into one block and all items into one block. If persons are assessed with multiple types of items, these items may be partitioned into blocks based on type, such as one block representing 2PL items and a different block representing 3-category generalized partial credit model (GPCM) items. Moreover, multiple person blocks may be used, for example, to maintain consistency of scale when calibrating items that may have been exposed to nonequivalent groups of test takers or for multi-group analyses allowed in various IRT calibration software \parencite[e.g.,][]{rpackage:mirt,chung2020}. The flexibility of block definitions for both persons and items provides support for a variety of psychometric analyses and data collection designs.

For clarity, we introduce the following notation for corresponding indices and counts. Quantities defined for persons (e.g., proficiency parameters) are indexed by $i$. Persons are partitioned into blocks, and the block index of person $i$ is denoted by $b(i) \in \{1,\ldots,B_{\I}\}$, where $B_{\I}$ is the total number of person blocks. The set of person indices for the persons in block $b$, for each $b \in \{1,\ldots,B_{\I}\}$, is denoted by $\mathcal{I}_{b} = \{i: b(i) = b\}$. Quantities defined for items (e.g., item parameters) are indexed by $j$. Items are partitioned into blocks, and the block index of item $j$ is denoted by $b(j) \in \{1,\ldots,B_{\J}\}$, where $B_{\J}$ is the total number of item blocks. The set of item indices for the items in block $b$, for each $b \in \{1,\ldots,B_{\J}\}$, is denoted by $\mathcal{J}_{b} = \{j: b(j) = b\}$. Item responses are indexed by $n$. Each item response results from the pairing of one person and one item. The person index associated with response $n$ is denoted by $i(n)$, and the item index associated with response $n$ is denoted by $j(n)$. The counts of persons, items, and responses in a particular dataset are denoted by $N_{\I}$, $N_{\J}$, and $N$, respectively, where $N_{\I}(b)$ and $N_{\J}(b)$ represent the number of persons in persons block $b$ and items in items block $b$, respectively.

\subsection{Latent Structure Models}

The vector of latent proficiency traits for person $i$ is denoted by $\boldsymbol{\theta}_i$. The dimension of $\boldsymbol{\theta}_i$ is assumed to be the same for all $i$, regardless of the associated person block, $b(i)$. This assumption supports common use cases for operations and simplifies estimation code, but the model theory and corresponding algorithm accommodate the case in which the dimension of $\boldsymbol{\theta}_i$ varies by block.

Indeed, for items, the dimension of the latent variables (i.e., item parameters with respect to IRT models) can and often must vary by block due to different items containing different grading mechanisms requiring IRT models with different numbers of parameters. Thus, the dimension of the latent variable for item $j$, $\boldsymbol{\psi}_j$, can vary as a function of the associated item block, $b(j)$. Items in the same block are assumed to be members of the same parametric family of IRT models and relate to the same person proficiency dimension(s). Between-item multidimensional IRT models \parencite{adams:etal:1997} can be specified via multiple item blocks, each of which relates to a single dimension of person proficiency. This structure also allows for within-item multidimensional IRT models, such that a given block represents a distinct combination of person proficiency dimensions rather than a single dimension.

Units in a given block may have a vector of observed predictors of their corresponding latent variables, hereafter referred to as `features'. To simplify notation, we generally use $\mathbf{X}$ to refer to such features, with the notation $\mathbf{X}_i$ referring to the vector of observed features for person $i$'s value of $\boldsymbol{\theta}_i$, and $\mathbf{X}_j$ referring to the vector of observed features for item $j$'s value of $\boldsymbol{\psi}_j$. The dimension of $\mathbf{X}$ depends on block but is constant for all units and latent variables within unit sharing a block.

The latent variable vectors in a block share a common probability model. Letting $\mathbf{u}$ denote a vector of latent variables for an arbitrary unit in an arbitrary block, we assume that
\begin{equation}
    \label{eq:ULR}
    \mathbf{u} = \mathbf{B}^{\prime}\mathbf{X} + \boldsymbol{\epsilon},
\end{equation}
where $\mathbf{u}$ is a vector with length denoted by $K$, $\mathbf{X}$ is a vector with length denoted by $p$ and represents the unit's observed features, $\mathbf{B}$ is a $(p \times K)$ matrix of regression parameters, and $\boldsymbol{\epsilon}$ is assumed to be multivariate normal with mean vector $\mathbf{0}$ and $(K \times K)$ covariance matrix $\boldsymbol{\Gamma}$. All cross-unit model parameters are assumed to be independent both within and across blocks.

Each block thus has its own values of $\mathbf{B}$ and $\boldsymbol{\Gamma}$, which specifies the probability distribution of the units in that block via the latent regression in~\eqref{eq:ULR}. We let $\boldsymbol{\Lambda}^{\I}_{b} = (\mathbf{B}^{\I}_{b},\boldsymbol{\Gamma}^{\I}_{b})$ denote the latent regression parameters for person block $b \in \{1,\ldots,B_{\I}\}$. Analogously, we let $\boldsymbol{\Lambda}^{\J}_{b} = (\mathbf{B}^{\J}_{b},\boldsymbol{\Gamma}^{\J}_{b})$ denote the latent regression parameters for item block $b \in \{1,\ldots,B_{\J}\}$. For item blocks, it may be desirable to transform item parameters to an unconstrained scale (e.g., logarithms of positive parameters, or logit transformations of bounded parameters), so that the regression specification in~\eqref{eq:ULR} is sensible given the proposed prior distributions described below.

\subsection{Item Response Models}

Let $Y_n$ denote the item response for observation $n$, which results from person $i(n)$ responding to item $j(n)$. We denote the probability model for $Y_n$ by $p(Y_n \mid \boldsymbol{\theta}_{i(n)}, \boldsymbol{\psi}_{j(n)})$, which is determined by the IRT model corresponding to item $j(n)$. Item responses are assumed to be conditionally independent of both person and item observed features. Moreover, item responses are assumed to be conditionally independent of one another, given the corresponding person and item latent variables. These are common assumptions in standard IRT modeling.

\subsection{Joint Probability Model}

Inferences for item and person latent variables, as well as the parameters governing their population distributions by block, are based on the following joint probability distribution for a collection of $N$ item responses from $N_{\I}$ persons and $N_{\J}$ items:
\begin{equation}
    \label{eq:jointmodel}
    \left[\prod_{n=1}^{N}p(Y_n \mid \boldsymbol{\theta}_{i(n)}, \boldsymbol{\psi}_{j(n)})\right]
    \left[\prod_{b=1}^{B_{\I}}p(\boldsymbol{\Lambda}^{\I}_{b}) \prod_{i \in \mathcal{I}_b}p(\boldsymbol{\theta}_i \mid \mathbf{X}_i, \boldsymbol{\Lambda}^{\I}_{b})\right]
    \left[\prod_{b=1}^{B_{\J}}p(\boldsymbol{\Lambda}^{\J}_{b}) \prod_{j \in \mathcal{J}_b}p(\boldsymbol{\psi}_j \mid \mathbf{X}_j, \boldsymbol{\Lambda}^{\J}_{b})\right].
\end{equation}
The distribution in~\eqref{eq:jointmodel} reflects the conditional independence assumptions previously noted. Note that $\{p(\boldsymbol{\Lambda}^{\I}_{b})\}_{b=1}^{B_{\I}}$ and $\{p(\boldsymbol{\Lambda}^{\J}_{b})\}_{b=1}^{B_{\J}}$ are the prior distributions for the latent regression parameters for person blocks and item blocks, respectively. Therefore, given the observed features and responses, the full posterior distribution of all parameters (persons, items, and latent regression) is proportional to~\eqref{eq:jointmodel}.

\subsection{Prior Distributions and Identification Constraints}\label{sec:prior}

All estimation of model parameters requires model identification. The need for identifying constraints results from the location and scale indeterminacy of persons and items in IRT models. A common set of constraints specify that $\mbox{E}[\boldsymbol{\theta}_i] = \mathbf{0}$ and the diagonal elements of $\boldsymbol{\Gamma}^{\I}_b$ equal 1 for an arbitrary person block $b$ (e.g., \cite{rpackage:mirt}). Provided that the observed pairs of persons and items are sufficiently ``connected'' across blocks \parencite{weeks:williams:1964}, such a constraint on the location and scale of the person parameters can ensure identification of the remaining latent regression parameters in~\eqref{eq:jointmodel}. However, in some applications (e.g., test equating with anchor items, \cite{kim2006}), constraints on the item side may be preferable. Our system supports fixing latent regression parameters for any number of blocks, and/or fixing unit parameters for any number of units, to support a variety of constraints including those required for identification.

One complication caused by identification constraints is difficulty in using standard conjugate prior distribution specifications for Bayesian multivariate linear regression. Letting $(\mathbf{B},\boldsymbol{\Gamma})$ denote the latent regression parameters for an arbitrary block, a matrix normal prior distribution for $\mathbf{B}$ and an inverse Wishart prior distribution for $\boldsymbol{\Gamma}$ would yield full conditional distributions of the same parametric families (e.g, \cite{Box:Tiao:1973}) that would simplify Gibbs sampling. The location constraints are straightforward to impose within this framework, but it is inconvenient to deal with a restriction to the inverse Wishart distribution, in which the random matrices are constrained to have fixed diagonal elements. In addition, there are known limitations of the inverse Wishart distribution as a weakly informative prior distribution for variance parameters \parencite{konig:etal:2020,Gelm:2006,Mari:McCa:Lock:2010}.

Thus, we adopt the ``separation'' strategy of \parencite{Barn:McCu:Meng:2000}, in which $\boldsymbol{\Gamma}$ is decomposed as $\mathbf{S} \mathbf{R} \mathbf{S}$ for diagonal matrix $\mathbf{S}$ of standard deviations and symmetric matrix $\mathbf{R}$ of correlations. Moreover, this decomposition allows for independent prior distributions to be used for $\mathbf{S}$ and $\mathbf{R}$. This separation enables straightforward imposition of scale identification constraints. For instance, the diagonal elements of $\mathbf{S}$ can be fixed for an arbitrary person block to identify the scale of the person parameters, whereas $\mathbf{R}$ for that block can be freely estimated.

We now specify the prior distribution for the latent regression parameters $\boldsymbol{\Lambda} = (\mathbf{B},\mathbf{S},\mathbf{R})$ for an arbitrary block (where $\boldsymbol{\Lambda}$ has been reparameterized to be in terms of $\mathbf{S}$ and $\mathbf{R}$ rather than $\boldsymbol{\Gamma}$), letting $K$ denote the dimension of the latent regression for the block and $p$ the number of feature variables for the block. Let $\mathbf{b} = \vect(\mathbf{B})$ be the $pK$ length vector of the stacked columns of the $(p \times K)$ matrix $\mathbf{B}$ defined in~\eqref{eq:ULR}. In the absence of constraints, the unknown parameters for this block are the $pK$ elements of $\mathbf{b}$, the $K$ diagonal elements of $\mathbf{S}$, and the $K(K-1)/2$ free parameters of $\mathbf{R}$. With no features, $p=1$, corresponding to a freely estimated intercept for each of the $K$ dimensions of the block's latent variables. The latent regression parameters are assumed to be independent across blocks.

We assume that $p(\mathbf{b},\mathbf{S},\mathbf{R}) = p(\mathbf{b})p(\mathbf{S})p(\mathbf{R})$, where $p(\mathbf{b})$ is a multivariate normal distribution with a specified mean vector $\mathbf{b}_0$ and specified precision matrix $\boldsymbol{\Omega}_0$; $p(\mathbf{S})$ is the product of $K$ independent uniform distributions with specified vectors of lower bounds $\mathbf{S}_{\ell}$ and upper bounds $\mathbf{S}_{u}$; and $p(\mathbf{R})$ is proportional to $|\mathbf{R}|^{\eta - 1}$ for a specified scalar parameter $\eta > 0$, and with $|\mathbf{R}|$ the determinant of $\mathbf{R}$. The latter corresponds to the LKJ distribution \parencite{lewandowski:etal:2009}, implemented in Stan as described in the \href{https://mc-stan.org/docs/functions-reference/lkj-correlation.html}{Stan documentation}. The LKJ distribution has support on the space of positive-definite correlation matrices. In our applications, we set $\eta = 1$, which yields the uniform distribution on the space of all correlation matrices as suggested by \textcite{Barn:McCu:Meng:2000}.

\section{MCMC Estimation Algorithm}

The MCMC estimation algorithm is implemented as a Gibbs sampler for the joint probability model specified in~\eqref{eq:jointmodel}. Gibbs sampling allows us to partition the full joint probability model into a set of conditional probability models and then iteratively sample from these conditional distributions \parencite{casella1992a, gelfand2000}. As described by \textcite{gelfand2000}, the ``form of each [conditional distribution] determines which [sampling] method is most suitable for a given [parameter]'' (p. 1301). Some conditional distributions might be sampled with a known distribution whereas others might require more general sampling methods. Specifically, for the most intractable conditional densities, one could update the parameters with a Metropolis-Hastings step, which is often referred to as ``Metropolis-Hastings within Gibbs'', although as the Gibbs sampler is a special case of the Metropolis-Hastings sampler, some statisticians have argued that this terminology is redundant \parencite[p. 332]{chib1995}.

In the following subsections, we will describe our strategies for sampling from these conditional densities.

\subsection{Unit Parameters}

The person and item parameter update steps are symmetrical. Each unit parameter within a block is updated using the Metropolis-Hastings algorithm. For instance, let $\mathcal{N}(i) = \{n: i(n)=i\}$ be the set of responses linked to person $i$. Then from~\eqref{eq:jointmodel}, the full conditional distribution for $\boldsymbol{\theta}_i$ is proportional to 
\begin{equation*}
    \left[\prod_{n \in \mathcal{N}(i)}p(Y_n \mid \boldsymbol{\theta}_{i}, \boldsymbol{\psi}_{j(n)})\right]
    p(\boldsymbol{\theta}_i \mid \mathbf{X}_i, \boldsymbol{\Lambda}^{\I}_{b(i)})
\end{equation*}
and is sampled using a Metropolis-Hastings step \parencite{chib1995}. Similarly, to update the parameters corresponding to item $j$, let $\mathcal{N}(j) = \{n: j(n)=j\}$ be the responses linked to item $j$. Then the full conditional distribution for $\boldsymbol{\psi}_j$ is proportional to
\begin{equation*}
    \left[\prod_{n \in \mathcal{N}(j)}p(Y_n \mid \boldsymbol{\theta}_{i(n)}, \boldsymbol{\psi}_{j})\right]
    p(\boldsymbol{\psi}_j \mid \mathbf{X}_j, \boldsymbol{\Lambda}^{\J}_{b(j)})
\end{equation*}
and is sampled using a Metropolis-Hastings step. For each of these steps, we assume that parameters corresponding to all other unit and latent regression parameters are fixed to samples from the previous step in the algorithm. Moreover, all units within a given block are independently updated. This assumption allows us to improve efficiency of the algorithm by sampling units in parallel. See Section~\ref{sec:implementation} for more information about how parallelization is implemented within the algorithm.

\subsection{Latent Regression Parameters}

As with the unit parameters, all latent regression structures (for all item and person blocks) are similar and updated with symmetric logic. We describe a general update procedure that applies to all blocks and is implemented sequentially at each iteration. For the remainder of this section, assume a specific block of either persons or items, eliminating the need for block or unit type indices.

From~\eqref{eq:jointmodel}, full conditional distributions for latent regression parameters in a single block are analogous to those for a Bayesian multivariate linear regression model with normally distributed residuals. Let $K$ be the dimension of the regression, $p$ the number of features in the regression, $U$ the number of units where $U = N_{\I}(b)$ for person block $b$ and $U = N_{\J}(b)$ for item block $b$, and $\mathbf{X}_0$ the $(U \times p)$ matrix of features common across the $K$ dimensions with the $u$th row corresponding to $\mathbf{X}_u^{\prime}$. This commonality simplifies the procedures due to results on seemingly unrelated regressions \parencite{Zell:1962,zellner:ando:2010}. Assume that $\mathbf{X}_0$ is of full column rank so that all parameters are identified and potentially includes a column of 1s to represent the intercept, depending on the identification constraints.

Let $\mathbf{U}$ denote the vector of unit parameters of length $UK$, ordered such that the first $U$ elements correspond to dimension 1, the next $U$ elements correspond to dimension 2, etc. Then the conditional distribution of $\mathbf{U}$ given $\mathbf{X}_0$ and the block's values of $\mathbf{b}$ and $\boldsymbol{\Gamma}$ is multivariate normal with mean vector $(\mathbf{I}_K \otimes \mathbf{X}_0)\mathbf{b}$ and covariance matrix $(\boldsymbol{\Gamma} \otimes \mathbf{I}_{\U})$, where $\mathbf{I}_m$ is the $(m \times m)$ identity matrix, $\boldsymbol{\Gamma} = \mathbf{S}\mathbf{R}\mathbf{S}$, and $\otimes$ is the Kronecker product. This result follows from the assumption of normally distributed residuals for the latent regressions in~\eqref{eq:jointmodel}.

The remainder of this section describes the update steps for $\mathbf{B}$, $\mathbf{R}$, and $\mathbf{S}$. As described earlier, when updating each of $\mathbf{B}$, $\mathbf{R}$, $\mathbf{S}$, or $\mathbf{U}$, all parameters save the one being updated are held constant due to the Gibbs algorithm.

\subsubsection{Updating \texorpdfstring{$\mathbf{B}$}{B}}

Elements of $\mathbf{b} = \vect(\mathbf{B})$ are jointly updated conditional on the current values of $\mathbf{U}$ and $\boldsymbol{\Gamma}$ using standard results for Bayesian multivariate linear regression \parencite[e.g.,][]{Lind:Smit:1972,zellner:ando:2010}. Recall that $\mathbf{b}$ has prior distribution $N(\mathbf{b}_0, \boldsymbol{\Omega}_0^{-1})$ where $\mathbf{b}_0$ is the prior mean vector and $\boldsymbol{\Omega}_0^{-1}$ is the prior precision matrix. Define
\begin{equation}
    \label{eq:bhat}
    \widehat{\mathbf{b}} = \left( \mathbf{I}_K \otimes (\mathbf{X}_0^{\prime}\mathbf{X}_0)^{-1}\mathbf{X}_0^{\prime} \right) \mathbf{U},
\end{equation}
the vector of OLS estimates of the regression coefficients stacked by dimension, equivalent to the generalized least squares (GLS) estimates in the case of seemingly unrelated regression \parencite{Zell:1962}. Next, define
\begin{equation}
    \label{eq:bpostprecision}
    \mathbf{P} = (\boldsymbol{\Gamma}^{-1} \otimes \mathbf{X}_0^{\prime}\mathbf{X}_0) + \boldsymbol{\Omega}_0.
\end{equation}
Then the full conditional distribution for $\mathbf{b}$ is multivariate normal with covariance matrix $\mathbf{P}^{-1}$ and mean vector
\begin{equation}
    \label{eq:bpostmean}
    \mathbf{P}^{-1}\left( (\boldsymbol{\Gamma}^{-1} \otimes \mathbf{X}_0^{\prime}\mathbf{X}_0) \widehat{\mathbf{b}} + \boldsymbol{\Omega}_0 \mathbf{b}_0 \right).
\end{equation}
An updated $\mathbf{b}$ is then sampled from this distribution.

\subsubsection{Updating \texorpdfstring{$\mathbf{R}$}{R}}

The correlation matrix $\mathbf{R}$ is updated by applying the Metropolis-Hastings algorithm to a transformation of the parameters of $\mathbf{R}$ to an unconstrained parameter space. The transformation we use is analogous to that used by \textcite{stan:2015} described in the \href{https://mc-stan.org/docs/reference-manual/transforms.html#cholesky-factors-of-correlation-matrices}{Stan manual}. The transformation has three steps. The first step transforms $\mathbf{R}$ to its unique lower-triangular Cholesky decomposition $\mathbf{L}$ satisfying $\mathbf{R} = \mathbf{L}\mathbf{L}^{\prime}$. The $K(K+1)/2$ non-zero elements of $\mathbf{L}$ are subject to $K$ quadratic constraints $\diag(\mathbf{L}\mathbf{L}^{\prime}) \equiv \mathbf{1}$, so that $\mathbf{L}$ has $K(K-1)/2$ free parameters. The second and third steps of the transformation of $\mathbf{R}$ reparameterize the free parameters of $\mathbf{L}$ to an unconstrained scale to simplify the proposal of candidate values. All three steps are detailed in the Appendix. The expressions presented in this section and the next reflect only the first step of the transformation (from $\mathbf{R}$ to $\mathbf{L}$) in order to simplify those expressions.

As shown in the Appendix, the transformation from $\mathbf{R}$ to $\mathbf{L}$ yields that when the prior distribution for $\mathbf{R}$ satisfies $p(\mathbf{R}) \propto |\mathbf{R}|^{\eta - 1}$, the prior distribution for $\mathbf{L}$ satisfies
\begin{eqnarray}
    \label{eq:pL}
     p(\mathbf{L}) & \propto & \prod_{k=2}^{K} \ell_{kk}^{K - k + 2\eta - 2},
\end{eqnarray}
where $\ell_{kk}$ is the $k$th diagonal element of $\mathbf{L}$. The full conditional distribution of $\mathbf{L}$ combines the prior distribution in~\eqref{eq:pL} with the likelihood contribution for $\mathbf{L}$ determined from the latent regression for the block. This contribution equals
    \begin{equation}
        |(\boldsymbol{\Gamma} \otimes \mathbf{I}_{\U})|^{-1/2} \exp\left(-\frac{1}{2} \mathbf{e}^{\prime} (\boldsymbol{\Gamma} \otimes \mathbf{I}_{\U})^{-1}\mathbf{e}\right), \label{eq:Glike1}
    \end{equation} 
where $\mathbf{e} = \mathbf{U} - (\mathbf{I}_K \otimes \mathbf{X}_0)\mathbf{b}$, a vector of length $UK$ consisting of the current values of the residuals from the latent regression. The full conditional distribution for $\mathbf{L}$ is proportional to the product of~\eqref{eq:pL}~and~\eqref{eq:Glike1}. Manipulations detailed in the Appendix indicate that the logarithm of this full conditional density is
\begin{equation}
    \sum_{k=2}^{K}(K - k + 2\eta - 2 - U)\log(\ell_{kk}) -U\sum_{k=1}^{K}\log(s_{kk}) - \frac{1}{2} \tr\left[ (\mathbf{E}'\mathbf{E}) (\mathbf{S}^{-1}\mathbf{L}^{-T}\mathbf{L}^{-1}\mathbf{S}^{-1}) \right], \label{eq:Llogfullcond}
\end{equation}
where $\mathbf{L}^{-T} \equiv (\mathbf{L}^{-1})^{\prime}$, $\mathbf{E}$ is the $(U \times K)$ matrix such that $\vect(\mathbf{E}) = \mathbf{e}$, and $s_{kk}$ is the $k$th diagonal element of $\mathbf{S}$. As previously noted, the $K(K-1)/2$ free parameters of $\mathbf{L}$ are updated indirectly via $K(K-1)/2$ parameters transformed to an unconstrained scale. The density of the unconstrained parameters includes a Jacobian correction for the transformation from the unconstrained scale to the constrained L, which is described in the Appendix. Finally, these transformed parameters are updated one element at a time using the Metropolis-Hastings algorithm. Note that $\boldsymbol{\Gamma}^{-1} = \mathbf{S}^{-1}\mathbf{R}^{-1}\mathbf{S}^{-1}=\mathbf{S}^{-1}\mathbf{L}^{-T}\mathbf{L}^{-1}\mathbf{S}^{-1}$ depends on every element of $\mathbf{L}$ and must be updated for each element at both the current and proposed values.

\subsubsection{Updating \texorpdfstring{$\mathbf{S}$}{S}}

Each diagonal element $s_{11}, \ldots, s_{KK}$ of $\mathbf{S}$ is updated one at a time, using the Metropolis-Hastings algorithm conditional on $\mathbf{B}$, $\mathbf{L}$, $\mathbf{U}$, and other values of $\mathbf{S}$.  Because $p(s_{kk})$ is uniform for each $k$, the logarithm of its full conditional density follows from~\eqref{eq:Llogfullcond} as
\begin{equation}
     -U\log(s_{kk}) - \frac{1}{2} \tr \left[ (\mathbf{E}'\mathbf{E}) (\mathbf{S}^{-1}\mathbf{L}^{-T}\mathbf{L}^{-1}\mathbf{S}^{-1}) \right].
     \label{eq:LlogS}
\end{equation}
Note that $\boldsymbol{\Gamma}^{-1} = \mathbf{S}^{-1}\mathbf{R}^{-1}\mathbf{S}^{-1}=\mathbf{S}^{-1}\mathbf{L}^{-T}\mathbf{L}^{-1}\mathbf{S}^{-1}$ depends on all diagonal elements of $\mathbf{S}$ and must be updated for each $s_{kk}$ at both the current and proposed values. Moreover, the update algorithm requires a method for proposing a candidate value constrained to a bounded interval, described in the Appendix.

\subsubsection{Adding Person Weights}

One might include an optional vector of person weights, $w_i \geq 0$, which are normalized globally so that the sum of the weights is equal to the number of persons with positive weights. If we define $U_{\text{eff}, b}$ as $U_{\text{eff}, b} = \sum_{i \in \mathcal{I}_b} w_i$ for block $b$, then $U_{b}$ (the number of units in block $b$) does not typically equal $U_{\text{eff}, b}$ (the effective number of units in block $b$ given weighting) unless all weights are the same.

These weights impact several objects that propagate through the updates. First, the likelihood contributions of individual persons are weighted by $w_i$ when updating item unit parameters. Second, the feature matrices are weighted by $\mathbf{W} = \diag(\{w_i : i \in \mathcal{I}_b\})$, so that $\mathbf{X}_0^{\prime}\mathbf{X}_0$ becomes $\mathbf{X}_0^{\prime}\mathbf{W}\mathbf{X}_0$ and $(\mathbf{X}_0^{\prime}\mathbf{X}_0)^{-1}\mathbf{X}_0^{\prime}$ becomes $(\mathbf{X}_0^{\prime}\mathbf{W}\mathbf{X}_0)^{-1}\mathbf{X}_0^{\prime}\mathbf{W}$ in all relevant equations. Third, the residual cross-product matrix $\mathbf{E}^{\prime}\mathbf{E}$ becomes $\mathbf{E}^{\prime}\mathbf{W}\mathbf{E}$. Finally, the $U$ in \eqref{eq:Llogfullcond} and \eqref{eq:LlogS} are now replaced by $U_{\text{eff}, b}$.

Note that the extension of SUR to the weighted case follows naturally from \textcite{Zell:1962} by making the appropriate substitutions in the original argument.

\subsection{Implementation Details}\label{sec:implementation}

All code was written in R and C with the main algorithm written in C using native C types. The R API to C uses the standard C interface provided in the R.h, Rinternals.h, RMath.h, and R\_ext/Rdynload.h header files (as described in the R Internals documentation, see \cite{rcoreteam2023}) and not alternative interfaces, such as those in the Rcpp \parencite{eddelbuettel2013} or cpp11 \parencite{vaughan2023} packages. The reason for using the standard interface rather than alternative extensions was to avoid some of the overhead required in maintaining the C++ objects (see the ``Motivations for cpp11 vignette'' in the cpp11 package, \cite{vaughan2023}). Moreover, the implementations from the R header files are generally included in a single C source file called ``API\_R.c'', the purpose of which is to process R objects, perform some rudimentary error checking, and build standard C structures required for the rest of the algorithm. All other C files mostly depend on typical C header files, such as math.h, stdlib.h, stdbool.h, and stdio.h, header files included in the source code, an assumed OpenMP \parencite{dagum1998} installation, and a separate OpenBlas \parencite{wang2013, xianyi2012} installation. This design allows relatively straightforward extensions to alternate APIs, such as those in Python \parencite{vanrossum2023}, several Javascript APIs depending on interpreters (see https://wiki.mozilla.org/ServerJS/C\_API), or native C/C++. These implementations have not yet been written but can be designed based on how these languages represent C objects. The MCMC portion of the algorithm is agnostic to the calling language, although the original wrapper algorithm was written in R. Note that several C files depend on math implementations in RMath.h for simplicity, although these might be rewritten to be self-contained in the future.

\subsubsection{Computational Efficiencies}

We adopted several strategies to optimize the storage and computational efficiencies of the algorithm. Most importantly, we represented objects with single rather than double precision. Although R stores numbers in double precision, the float package was written to store standard R objects in single precision \parencite{schmidt2017, schmidt2023}. Single precision objects typically take approximately half the storage space and are twice as fast in calculations. Even though computers are currently designed to represent objects in double precision \parencite[e.g.,][]{cockcroft2001}, in our experience, we have found that these performance gains persist. For instance, the original version of the algorithm ran approximately twice as fast on the developers' computers when we switched the storage precision to single precision.

To enable single precision floats, we included single precision versions of BLAS and LAPACK algorithms, as implemented in OpenBlas \parencite{wang2013}. R typically comes with and is built against BLAS and LAPACK with only double precision routines, although it is possible to compile R libraries against alternate implementations. BLAS and LAPACK include fast routines for standard matrix manipulations, such as matrix multiplication, inversion, and decompositions \parencite{blackford2002, anderson1999}. For additional performance and storage gains, we used packed (non-redundant) representations of triangular and symmetric matrices wherever possible unless they noticeably impacted performance, as in the case with Cholesky decomposition \parencite[e.g.,][]{gustavson2010}. Moreover, highly sparse matrices were always represented as packed arrays for processing and storage efficiency.

Sparsity of the person-item linkages was accommodated by creating a nested data structure for each block that stores the relevant linkage information for that block's units. For example, for a given person block, the data structure contains the observation, item block, and item unit indices of the item responses for each person in that block. Each item block stores the analogous information for how each of that block's items links to observations and persons. Encoding the linkage information into separate item and person views, and accessing that information via pointers, supports efficient looping required for Metropolis-Hastings updates of the unit parameters.

Finally, we employed OpenMP to parallelize updates over person and item units \parencite{dagum1998}. In general, MCMC algorithms are inherently sequential, as each iteration depends on the prior iteration. However, one can still exploit conditional independence within a Gibbs update. In our case, unit parameters within a block (either persons or items) are updated independently given the current values of the other parameters. Rather than partitioning data across chains and aggregating samples of parameters \parencite[e.g.,][]{neiswanger2014}, we distributed person and item-level updates with OpenMP across threads. OpenMP's shared-memory model \parencite{chandra2001a} distributes work concurrently across threads without duplicating large data structures. As all threads operate on a common state, this strategy provided a mechanism for within-block updates to run concurrently without risking storage/memory issues nor obfuscating implementation logic.

\subsubsection{Algorithm Tuning}

The Metropolis-Hastings algorithm relies on tuning parameters that determine how efficiently the joint posterior distribution of the parameters is sampled. In particular, one must specify the standard deviations required to propose random candidate parameter values.

We implemented a four-phase adaptive procedure to set these parameters. Iterations from only Phase 4 are used for inference, whereas Phases 1-3 comprise burn-in and are reserved for algorithm tuning. Phase 1 randomly initializes all parameters and runs the algorithm for a specified number of iterations ($M_1$) using relatively large proposal standard deviations for all parameters to encourage exploration.

Phase 2 runs the algorithm for a specified number of iterations using these same proposal standard deviations ($M_2$), and calculates the acceptance rate for each sampled parameter. The acceptance rates from Phase 2 are then used to adjust the proposal standard deviations for Phase 3. For parameters with acceptance rate below a specified lower-bound rate $a_{0}$, the proposal standard deviation is decreased by a factor of 5; for parameters with acceptance rate above a specified upper-bound rate $a_{1}$, the proposal standard deviation is increased by a factor of 5; and for parameters with acceptance rate between $a_{0}$ and $a_{1}$, the proposal standard deviation is unchanged.

Phase 3 then proceeds for a specified number of iterations ($M_3$) using these revised proposal standard deviations, and stores the acceptance rate for each parameter. For each parameter with a different proposal standard deviation for Phases 2 and 3, a function of these proposal standard deviations and the corresponding acceptance rates is used to compute a proposal standard deviation for Phase 4 that aims to achieve a specified target acceptance rate $a_{*}$ satisfying $a_0 < a_{*} < a_{1}$.

Finally, Phase 4 is run for a specified number of iterations ($M_4)$, with the acceptance rate, first moment, and second moment of each parameter stored for algorithm diagnostics, model diagnostics, and posterior inferences. Samples are optionally thinned by a specified amount for storage. Stored samples are used to compute Gelman-Rubin statistics \parencite{Gelm:Rubi:1992} to assess convergence across parallel chains, as well as to implement posterior predictive checks \parencite{Gelm:Meng:Ster:1996,sinharay:johnson:stern:2006} tailored to evaluating fit of IRT models. All samples from Phase 4 (whether stored or not) are also used to compute estimates of the Expected Log Pointwise Predictive Density (ELPD) and the Widely Applicable Information Criterion (WAIC) of \textcite{vehtari:etal:2017} for use in model comparisons.

\subsubsection{Extensibility}

The original implementation included several unidimensional IRT models: the 1PL \parencite{rasch1960}, 2PL and 3PL \parencite{birnbaum1958}, 4PL \parencite{loken2010}, GPCM \parencite{muraki1997}, continuous (Gaussian) response model \parencite[e.g.,][]{moustaki2000}, and multiple bounded-continuous response models \parencite[e.g.,][]{molenaar2022, zopluoglu2024}. All current implemented item types are unidimensional. Extending the algorithm to within-item multidimensionality would primarily require modifying the likelihood macros to accept arrays of item discrimination parameters rather than assuming that those parameters are scalars.

Adding new IRT models is similarly straightforward and modular: define the model and update the model index (and corresponding array) to include this model; specify the number of parameters; update the switch statement to pass item parameters, person parameters, and response vectors to the appropriate log-likelihood implementation; and implement the log-likelihood calculation. Note that the likelihood calculation need only include terms that depend on model (person and item) parameters, as constants cancel in the Metropolis-Hastings ratio. Minor updates are also required in the R API for the purpose of checking and updating response data, specifying model parameters, and performing post-estimation checks. As we prioritized model extensibility, we ensured that additional IRT models could be added with relative ease, simply by indicating when to apply the model and how to determine the log-likelihood.

% \section{Demonstration}
% \rednote{TBD}

\section{Discussion}

The model and methods described herein are useful for performing psychometric analyses of large, sparse item response datasets. These methods are particularly well-suited to IRT analysis of large banks of automatically generated items, for which a common assumption is of stochastic item parameters that are informed by a combination of observable item features and response data. Posterior distributions of item parameters can then be used to support both adaptive assignment algorithms and inferences about person proficiency \parencite[e.g.,][]{nydick2026, sharpnack2026}.

This software was developed specifically for estimating item parameter distributions on the Duolingo English Test \parencite[DET;][]{naismith2025}. A typical calibration cycle contains tens of millions of responses from hundreds of thousands of units. Estimating IRT models at this scale presents substantial computational and memory challenges. Existing software packages, such as \texttt{mirt} \parencite{rpackage:mirt} and Stan \parencite{stan:2015}, provide flexible frameworks for latent variable modeling but were not designed for this combination of response volume, custom model structure, and production-oriented calibration requirements. For example, calibrating items on the DET requires support for partitioning items and persons into blocks, incorporating item-level features, and adding several custom IRT models while also managing memory usage and minimizing runtime overhead. Accommodating these requirements within existing software would have involved substantial customization and computational tradeoffs that were impractical for routine use. We piloted existing software for large-scale calibration; however, at the scale of typical DET calibrations, many of these approaches either exhausted available memory or required runtimes on the order of days rather than hours.

One decision that reflected the production-oriented goal was the use of Metropolis-Hastings within Gibbs sampling rather than other methods, such as Hamiltonian Monte Carlo \parencite[HMC;][]{neal2026}. HMC produces accurate posterior samples with low rejection rates \parencite{betancourt2018,neal2026} but can work less well in large datasets or require substantial computational demands \parencite{thiagarajan2025a}. Moreover, standard HMC relies on continuous gradients and would require strategies to sample from distributions with discontinuities in the target distributions \parencite{nishimura2020}, limiting the complexity of allowed IRT models. Metropolis-Hastings-within-Gibbs algorithms avoid this constraint, scale straightforwardly as IRT model complexity increases \parencite[e.g.,][]{patz:junker:1999}, and exploit the natural block structure of IRT response processes: as described earlier, for a given Metropolis-Hastings step, persons or items within blocks can be updated in parallel. However, alternative strategies for MCMC sampling are discussed below as avenues for future work.

The model and algorithm design attempts to balance simplicity and flexibility. The use of linear latent regressions of a common parametric structure across all blocks supports only a subset of the kinds of models that one may want to fit to sparsely-linked item response data. But the restriction also allows efficient coding that supports scalability to large datasets and easy extensibility to additional, parameterized IRT models. Moreover, the ability to flexibly stratify both persons and items into subpopulations via block memberships allows corresponding software to support a broad set of psychometrically useful analyses. For example, multiple person and item blocks can be used to conduct item parameter linkage analysis in which nonequivalent person groups are linked by common items. Analysis of measurement non-invariance could then be conducted conditional on a set of assumed anchor items and a specified stratification of the person population based on observed person features.

More generally, blocks can be used to stratify any population of units into subpopulations based on observed features. Thus, the marginal latent distribution in the population can be expressed as a mixture distribution across strata defined by functions of unit features. This ability provides a practical method for addressing potential violations of the assumption of a linear latent regression with homoskedastic Gaussian residuals for the unit (item or person) population. For example, nonlinear relationships between features and latent variables can be approximated by stratification. It is also worth noting that the predictors used in linear latent regressions could be based on nonlinear functions of the initial features designed to predict the latent variables (e.g., item parameters) based on external criteria or data such as the output of a tuned machine learning model \parencite[e.g.,][]{li:etal:2025,peters:etal:2025}. These capabilities allow for separation of the feature generation and tuning process and the difficulty modeling process.

Many of the components of this algorithm draw on well-established methods: Metropolis-Hastings-within-Gibbs sampling has a long history in IRT \parencite{patz:junker:1999}; LKJ priors with Cholesky parameterizations are a standard approach to sampling covariance structures and implemented in Stan \parencite{stan:2015}; and explanatory IRT models with person and item features have been widely studied \parencite[e.g.,][]{deboeck:2008}. The main contribution of this paper is the integration of these methods into a coherent, scalable system for regularly calibrating sparse assessment data. Specifically, the system described herein contains symmetric person- and item-side latent regressions, flexible block structures, and multiple IRT response models with easy extensibility and within a C implementation that uses single-precision storage, sparse linkage structures, and targeted parallelization. Together, these design choices make Bayesian calibration feasible at scale and for routine operational use.

One limitation of the software as written is lack of a hierarchical random effects structure for units. For instance, items might be nested within passages. In this case, unit parameters would be drawn from a distribution modeled by hyperparameters that depend on nesting structure. The extension of the current structure and computational methods to accommodate hierarchical unit populations is a promising area for future work. Another area of future work is the implementation of other methods for sampling from the joint posterior distribution, such as Hamiltonian Monte Carlo methods \parencite{neal2026} as well as custom prior distributions for model parameters based on results from an earlier calibration. The implementation of alternative sampling procedures or adjustments to the prior distributions could take advantage of the data structures and storage of relevant quantities that have already been developed.

The design and functionality of this software facilitates integrating item calibration into a comprehensive administration and scoring system. This system includes methods for predicting item and person parameters from features. Given the DET automated item generation process, these methods allow items to enter calibration with informative prior distributions derived mostly from their features and before substantial response data have been collected. Moreover, Bayesian calibration returns posterior samples of item parameters that can be used in administration and scoring procedures that accommodate parameter uncertainty \parencite{sharpnack2026}. These scoring and administration algorithms enable productive piloting: items can be included as scored based solely on their feature-predicted parameters, reducing the bottleneck whereby unscored pilot items compete for limited test space and risk lowering reliability. Scoring pilot items with limited data can be critical in processing the large number of items that automated item generation processes can produce. Finally, the efficiencies described earlier enable maintaining assessments with large, dynamic item pools, sparse response matrices, and continuous calibration processes. Because this software returns posterior samples of item parameters, item parameter uncertainty propagates naturally into operational procedures; moreover, as per-item sample sizes grow, this posterior uncertainty is appropriately reduced given accumulating response data. In this way, SPICE supports an efficient, continuous, and self-updating calibration process.

The capabilities enabled by SPICE represent systems that are difficult to achieve at scale with existing software. The combination of an efficient sampling algorithm, symmetric explanatory structure, mixed and extensible IRT models, and full posterior uncertainty distinguishes this software from currently available alternatives and motivated its development. Moreover, any large-scale assessment sharing these properties (automated item generation, adaptive administration, sparse person-item linkages, and a need to score items with little to no response data) benefits from the same architectural choices described earlier. As AI-based item generation and next-generation adaptive assessment continue to expand the scale, complexity, and personalization of operational testing, algorithms of this kind are critical to ensuring that the scores constructed from these assessments remain statistically and psychometrically principled and that score interpretations are sensible regardless of the scale at which they are produced.

\newpage
\printbibliography 

\newpage
\appendix

% display only for arXiv format
\ifarxiv
  \appendixpage
\fi

\section{Density of Cholesky Decomposition under LKJ prior}\label{app:choldist}

Let $\mathbf{R}$ be a positive-definite correlation matrix of dimension $K$. The LKJ prior for $\mathbf{R}$ specifies that $p_{R}(\mathbf{R}) \propto |\mathbf{R}|^{\eta - 1}$ for a scalar parameter $\eta > 0$. Let $\mathbf{L}$ be the lower-triangular Cholesky decomposition of $\mathbf{R}$, satisfying $\mathbf{L}\mathbf{L}^{\prime} = \mathbf{R}$, and define the function $\mathbf{g}(\mathbf{R}) = \mathbf{L}$.  This function is continuous and invertible by standard results on the uniqueness of the Cholesky decomposition. The inverse function is $\mathbf{g}^{-1}(\mathbf{L}) = \mathbf{R} = \mathbf{L}\mathbf{L}^{\prime}$.  From standard results on transformations of random vectors, 
\begin{eqnarray}
    p_{L}(\mathbf{L}) & = & p_{R}(\mathbf{g}^{-1}(\mathbf{L})) \big\vert \frac{\partial}{\partial \mathbf{L}} \mathbf{g}^{-1}(\mathbf{L}) \big\vert \notag\\
    & \propto & | \mathbf{L} |^{2(\eta - 1)} \big\vert \frac{\partial}{\partial \mathbf{L}} \mathbf{L}\mathbf{L}^{\prime} \big\vert \label{appeq:pL},
\end{eqnarray}
where the second line follows from the fact that $|\mathbf{L}\mathbf{L}^{\prime}| = |\mathbf{L}|^2$. To evaluate the determinant of the Jacobian $\big\vert \frac{\partial}{\partial \mathbf{L}} \mathbf{L}\mathbf{L}^{\prime} \big\vert$, note that dimension of the parameter space for both $\mathbf{R}$ and $\mathbf{L}$ is $K(K-1)/2$, due to the $K$ quadratic constraints $\diag(\mathbf{L}\mathbf{L}^{\prime}) \equiv 1$. Organize the $K(K-1)/2$ free parameters of $\mathbf{R}$ as $\mathbf{r} = (r_{21},r_{31},r_{32},\ldots,r_{K1},\ldots,r_{K,K-1})^{\prime},$ obtained by reading across the rows of the lower triangle of $\mathbf{R}$.  Analogously organize the $K(K-1)/2$ free parameters of $\mathbf{L}$ as $\boldsymbol{\ell} = (\ell_{21},\ell_{31},\ell_{32},\ldots,\ell_{K1},\ldots,\ell_{K,K-1})^{\prime}.$ Note that each element $r_{ij}$ of $\mathbf{r}$ is the dot product of row $i$ and row $j$ of $\mathbf{L}$. Because $i > j$ and $\mathbf{L}$ is lower triangular, $r_{ij}$ is a function of $\ell_{ij}$ and other elements of $\boldsymbol{\ell}$ that occur prior to $\ell_{ij}$. Thus, the Jacobian is a lower triangular matrix, so its determinant is the product of the diagonal elements,
\begin{equation}
\prod_{j < i}\frac{\partial r_{ij}}{\partial \ell_{ij}},
\label{eq:partialRL}
\end{equation}
where $r_{ij} = \sum_{k = 1}^{j}\ell_{ik}\ell_{jk}$, so that $\frac{\partial r_{ij}}{\partial \ell_{ij}} = \ell_{jj}$. Each element of the product in~\eqref{eq:partialRL} is equal to one of the diagonal elements of $\mathbf{L}$, denoted by $\ell_{11},\ell_{22},\ldots,\ell_{KK}$. For example, the partial derivative of $r_{42}$ with respect to $\ell_{42}$ is $\ell_{22}$. Manual inspection of the $K(K-1)/2$ elements of the product in~\eqref{eq:partialRL} yields that $\ell_{11} = 1$ occurs $(K-1)$ times, $\ell_{22}$ occurs $(K-2)$ times, etc, until $\ell_{K-1,K-1}$ occurs $K-(K-1) = 1$ time.  Thus~\eqref{eq:partialRL} equals $\prod_{k=1}^{K-1} \ell_{kk}^{K-k} = \prod_{k=2}^{K} \ell_{kk}^{K-k}$, which is equivalent to \textcite{henderson1979} Equation (55) assuming constrained diagonals. Continuing the expression of~\eqref{appeq:pL} then yields
\begin{eqnarray*}
p_{L}(\mathbf{L}) & \propto & | \mathbf{L} |^{2(\eta - 1)} \prod_{k=2}^{K} \ell_{kk}^{K-k}\\
& = & \left( \prod_{k=2}^{K} \ell_{kk} \right)^{2(\eta - 1)} \prod_{k=2}^{K} \ell_{kk}^{K-k}\\
& = & \prod_{k=2}^{K} \ell_{kk}^{K - k + 2\eta - 2}.
\end{eqnarray*}

\section{Jacobian of Transformation Cholesky to Unconstrained Parameters}

To facilitate updating $\mathbf{L}$, it is convenient to transform its free parameters to an unconstrained scale. The two-step transformation we use is analogous to that used in Stan \parencite{stan:2015}, and the derivations included here are analogous to those in the \href{https://mc-stan.org/docs/reference-manual/transforms.html#variable-transforms.chapter}{Stan manual}. In the first step of the transformation, each of the free parameters of $\mathbf{L}$ is mapped to the open interval $(-1,1)$. In the second step, each of these values is mapped to the real line. Here we detail the transformations and their corresponding Jacobian determinants.

Organize the $K(K-1)/2$ free parameters of $\mathbf{L}$ as $\boldsymbol{\ell} = (\ell_{21},\ell_{31},\ell_{32},\ldots,\ell_{K1},\ldots,\ell_{K,K-1})^{\prime}$, and denote the diagonal elements of $\mathbf{L}$ by $\ell_{11},\ldots,\ell_{KK}$. For $2 \leq i \leq K$ and $j < i$ define
\begin{equation}
    z_{ij} = \frac{\ell_{ij}}{\sqrt{1 - \sum_{j' < j}\ell_{ij'}^2}},
    \label{eq:zij}
\end{equation}
where $\sum_{j' < j}\ell_{ij'}^2 = 0$ when $j=1$. The Euclidean length of each row of $\mathbf{L}$ is 1, implying that $-1 < z_{ij} < 1$. Define $\mathbf{z} = (z_{21},z_{31},z_{32},\ldots,z_{K1},\ldots,z_{K,K-1})^{\prime}$. The mapping $\boldsymbol{\ell} \to \mathbf{z}$ is the first step of the transformation. The second step is $\mathbf{z} \to \mathbf{y}$ where $y_{ij} = \tanh^{-1}(z_{ij})$. Thus the sequence $\boldsymbol{\ell} \to \mathbf{z} \to \mathbf{y}$ maps the $K(K-1)/2$ constrained parameters $\boldsymbol{\ell}$ to $K(K-1)/2$ unconstrained parameters $\mathbf{y}$.

Using this transformation to update $\mathbf{L}$ requires evaluation of the full conditional density of $\mathbf{y}$, with depends on the product of the prior density of $\mathbf{L}$ in~\eqref{eq:pL} and the determinants of the Jacobian matrices corresponding to the two transformation steps. The $\mathbf{y} \to \mathbf{z}$ transformation has a diagonal Jacobian matrix because $z_{ij} = \tanh(y_{ij})$. Thus the determinant is
\begin{equation}
    \prod_{j < i} \frac{1}{(\cosh(y_{ij}))^2},
    \label{eq:J1}
\end{equation}
where the product is taken over $(i,j)$ pairs with $j < i$.

Arguments analogous to those made in Appendix \ref{app:choldist} ensure that the Jacobian matrix of the $\mathbf{z} \to \boldsymbol{\ell}$ transformation is lower triangular. Thus the determinant is the product of the partial derivative of $\ell_{ij}$ with respect to $z_{ij}$ with $j < i$. Rearranging~\eqref{eq:zij} and taking the product of partial derivatives with $j < i$, we end up with a determinant of
\begin{equation}
    \prod_{j < i} \sqrt{ \left(1 - \sum_{j' < j}\ell_{ij'}^2 \right) }.
    \label{eq:J2}
\end{equation}

\section{Derivation of Log of Full Conditional Distribution of Cholesky}

Let $\mathbf{E}$ denote the $(U \times K)$ matrix such that $\vect(\mathbf{E}) = \mathbf{e}$.  Then $\mathbf{e}^{\prime} (\boldsymbol{\Gamma} \otimes \mathbf{I}_{\U})^{-1}\mathbf{e}$ equals $\vect(\mathbf{E})^{\prime} (\boldsymbol{\Gamma}^{-1} \otimes \mathbf{I}_{\U}) \vect(\mathbf{E})$. Note that $(\boldsymbol{\Gamma}^{-1} \otimes \mathbf{I}_{\U}) \vect(\mathbf{E})$ equals $\vect( \mathbf{I}_{\U} \mathbf{E} \boldsymbol{\Gamma}^{-1} )$ from standard results relating Kronecker products and the $\vect$ operator (e.g, \cite[p. 74]{Gent:2007}). Thus, $\mathbf{e}^{\prime} (\boldsymbol{\Gamma} \otimes \mathbf{I}_{\U})^{-1}\mathbf{e}$ equals $\vect(\mathbf{E})^{\prime} \vect( \mathbf{E} \boldsymbol{\Gamma}^{-1}) = \tr(\mathbf{E}^{\prime} \mathbf{E} \boldsymbol{\Gamma}^{-1})$. Starting with \eqref{eq:Glike1}, we have
    \begin{equation}
        |(\boldsymbol{\Gamma} \otimes \mathbf{I}_{\U})|^{-1/2} \exp\left(-\frac{1}{2} \mathbf{e}^{\prime} (\boldsymbol{\Gamma} \otimes \mathbf{I}_{\U})^{-1}\mathbf{e}\right) =
        |\boldsymbol{\Gamma}|^{-U/2} \exp\left(-\frac{1}{2} \tr (\mathbf{E}'\mathbf{E} \boldsymbol{\Gamma}^{-1})  \right)
        \label{eq:Glike2}.
    \end{equation}
The full conditional distribution of $\mathbf{L}$ is proportional to the product of \eqref{eq:pL}~and~\eqref{eq:Glike2}. We wish to show that the logarithm of this product is~\eqref{eq:Llogfullcond}. Note that $\boldsymbol{\Gamma} = \mathbf{S}\mathbf{L}\mathbf{L}^{\prime}\mathbf{S}$, so that $\boldsymbol{\Gamma}^{-1} = \mathbf{S}^{-1}\mathbf{L}^{-T}\mathbf{L}^{-1}\mathbf{S}^{-1}$.  Also, by standard properties of determinants, $|\boldsymbol{\Gamma}| = |\mathbf{L}|^{2}|\mathbf{S}|^{2}$.  Thus the product of~\eqref{eq:pL}~and~\eqref{eq:Glike2} equals
\begin{equation}
    \left( \prod_{k=2}^{K} \ell_{kk}^{K - k + 2\eta - 2} \right) |\mathbf{L}|^{-U} |\mathbf{S}|^{-U} \exp\left(-\frac{1}{2} \tr \left[ (\mathbf{E}'\mathbf{E}) (\mathbf{S}^{-1}\mathbf{L}^{-T}\mathbf{L}^{-1}\mathbf{S}^{-1}) \right]  \right). \label{eq:Lfullcond} 
\end{equation}
Each of $\mathbf{L}$ and $\mathbf{S}$ is triangular, and furthermore $\ell_{11} = 1$, so that $|\mathbf{L}| = \prod_{k=2}^{K}\ell_{kk}$ and $|\mathbf{S}| = \prod_{k=1}^{K}s_{kk}$. Thus the logarithm of~\eqref{eq:Lfullcond} yields~\eqref{eq:Llogfullcond}.

\subsection{Metropolis-Hastings Proposal for a Bounded Parameter}

This section specifies a method for proposing a candidate value for a scalar parameter $Y$ restricted to a bounded interval $(a,b)$, used for updating the elements of $\mathbf{S}$ for a block. Suppose a given scalar parameter has current value $y_0$ in the MCMC algorithm, and we wish to propose a candidate value $y_1$ restricted to the interval $(a,b)$. We generate $X \sim N(\logit(\frac{y_0 - a}{b-a}), \sigma)$ and set $Y_1 = a + (b-a)\logit^{-1}(X)$, where $\logit(p) = \log{[p/(1-p)]}$ and $\logit^{-1}(x)$ is the inverse logit transformation $\exp(x)/(1 + \exp(x))$.  The value $\sigma$ is a specified tuning parameter of the algorithm. To apply the M-H algorithm with this proposal distribution, we need the density ratio $$\frac{q(y_0 \mid y_1)}{q(y_1 \mid y_0)}.$$
From standard results on monotonic transformations of scalar random variables
\begin{eqnarray}
    q(y_1 \mid y_0) & = & \frac{1}{\sigma \sqrt{2\pi}} \exp\left( -\frac{1}{2\sigma^2}\Big(\logit\big((y_1-a)/(b-a)\big) - \logit\big((y_0-a)/(b-a)\big)\Big)^2\right) \nonumber\\
    & \times & \left(\frac{d}{dy} \logit\big((y - a) / (b - a)\big)\Big|_{y = y_1}\right)\nonumber\\
    & = & \frac{1}{\sigma \sqrt{2\pi}} \exp\left( -\frac{1}{2\sigma^2}\Big(\logit\big((y_1-a)/(b-a)\big) - \logit\big((y_0-a)/(b-a)\big)\Big)^2\right) \nonumber\\
    & \times & \frac{(b - a)}{(y_1 - a)(b - y_1)}
    \label{eq:mhbounded}
\end{eqnarray}
All but the last term of \eqref{eq:mhbounded} cancel in the ratio due to symmetry, so
\begin{equation}
    \frac{q(y_0 \mid y_1)}{q(y_1 \mid y_0)} =\frac{(y_1 - a)(b - y_1)}{(y_0 - a)(b - y_0)}.
    \label{eq:mhboundedratio}
\end{equation}

\end{document}